\documentclass[aps,prl,twocolumn,floatfix,showpacs]{revtex4}
\usepackage{graphicx}
\usepackage{dcolumn}
\usepackage{bm}
\begin{document}
\title{Raman and Infrared-Active Phonons in Hexagonal HoMnO$_3$
    Single Crystals: Magnetic Ordering Effects}
\author{A.P.~Litvinchuk,$^1$ M.N.~Iliev,$^1$ V.N. Popov,$^2$ and M.M. Gospodinov$^3$ }
\affiliation{$^1$Texas Center for Superconductivity and Advanced Materials, and\\
Department of Physics, University of Houston, Houston, Texas
77204-5002}
\affiliation{$^2$Faculty of Physics, University of
Sofia, 1164 Sofia, Bulgaria}
\affiliation{$^3$Institute of Solid
State Physics, Bulgarian Academy of Sciences, 1184 Sofia,
Bulgaria}
\date{September 8, 2003}

\begin{abstract}
Polarized Raman scattering and infrared reflection spectra of
hexagonal HoMnO$_3$ single crystals in the temperature range
10-300~K are reported. Group-theoretical analysis is performed and
scattering selection rules for the second order scattering
processes are presented. Based on the results of lattice dynamics
calculations, performed within the shell model, the observed lines
in the spectra are assigned to definite lattice vibrations. The
magnetic ordering of Mn ions, which occurs below T$_N$=76~K, is
shown to effect both Raman- and infrared-active phonons, which
modulate Mn-O-Mn bonds and, consequently, Mn exchange interaction.
\end{abstract}

\maketitle

\section{Introduction}
The hexagonal $R$MnO$_3$ [$R$ = Ho, Er, Tm, Yb, Lu, Y; space group
$P6_3cm (C^3_{6v}), Z=6$] compounds belong to the class of
ferroelectromagnet materials characterized by the coexistence of
antiferromagnetic (AFM) and ferroelectric (FE) orderings
\cite{smolenskii1}. The FE and AFM transitions are well separated
in temperature with Curie and N\'eel temperatures being $T_C >
800$~K and $T_N \approx 76$~K, and only weak coupling of the two
respective order parameters is expected. Such coupling do exist in
$R$MnO$_3$ materials. For YMnO$_3$, for instance, anomalies in the
dielectric constant and loss tangent near $T_N$ \cite{huang1} and
an additional antiferromagnetic contribution to the non-linear
optical polarizability below $T_N$ \cite{frohlich1,sakano1,fiebig}
were found experimentally. Furthermore, a number of magnetic
transitions at lower temperatures (below T$_N$) were established
for HoMnO$_3$ due to both Mn in-$xy$-plane and Ho $z$-axis
ordering\cite{fieb00,munoz1}. The Raman scattering is also
determined by non-linear terms of polarizability, which may be
affected by AFM-FE and/or spin-phonon couplings. In the case of
strong couplings the Raman spectra should exhibit anomalies at
magnetic transition temperatures. Anomalous Raman scattering due
to two-magnon processes was recently reported for
YMnO$_3$\cite{takahashi1}, but not confirmed in more recent
studies \cite{iliev1,sato}. All theses facts give a motivation for
a more thorough study of lattice vibrations of hexagonal
manganites as a function of temperature through Raman scattering
and infrared spectroscopies.

In this paper we report the polarized Raman and infrared
reflection spectra of HoMnO$_3$ single crystals in a broad
temperature range. Pronounced phonon anomalies, that are related
to spin-phonon or AFM-FE couplings, are observed experimentally. A
comparison to the mode frequencies predicted by lattice dynamical
calculations (LDA) allowed assignment of the Raman and infrared
lines to definite phonon modes or two-phonon Raman scattering
processes. No evidence for two-magnon scattering was found in the
low temperature antiferromagnetic phase.

\section{Samples and Experimental}
Pure polycrystalline hexagonal HoMnO$_3$ was synthesized by a
solid-state reaction of stoichiometric amounts of Ho$_2$O$_3$
(99.99\%) and MnO$_2$ (99.99\%), and further annealed for 24h at
1120$^\circ$C in oxygen atmosphere. HoMnO$_3$ single crystals were
grown by High Temperature Solution Growth Method using
PbF$_2$/PbO/B$_2$O$_3$ flux (PbF$_2$ : PbO : B$_2$O$_3$ = 0.8 :
0.195 : 0.005). The flux was mixed with HoMnO$_3$ powder in a 7 :
1 ratio) and annealed in a platinum crucible at 1250$^\circ$C for
48h in oxygen. After that the temperature was decreased down to
1000$^\circ$C at a rate of 0.5~C/h. The flux was decanted and
well-shaped hexagonal plate-like crystals of typical size $3
\times 5 \times 0.2$~mm removed from the bottom of the crucible.

The Raman spectra were measured under a microscope using a HR640
spectrometer equipped with a liquid-nitrogen-cooled CCD detector.
The 514.5~nm and 632.8~nm lines of Ar$^+$ and He-Ne laser were
used for excitation. The infrared reflectance was measured with a
Bomem DA8 Fourier-Transform interferometer equipped with a
near-normal incidence reflectance stage and a liquid-helium-cooled
bolometer. The real and imaginary parts of dielectric function,
$\epsilon_1(\omega)$ and $\epsilon_2(\omega)$, were obtained
through Kramers-Kronig analysis.

The lattice dynamical calculations for HoMnO$_3$ were performed
within the shell model used earlier for YMnO$_3$\cite{iliev2}. The
lattice parameters and atomic positions were taken from
Ref.\cite{munoz1}.

\section{Results and Discussion}
\subsection{Room temperature data}

\begin{figure}
\includegraphics[width=3in]{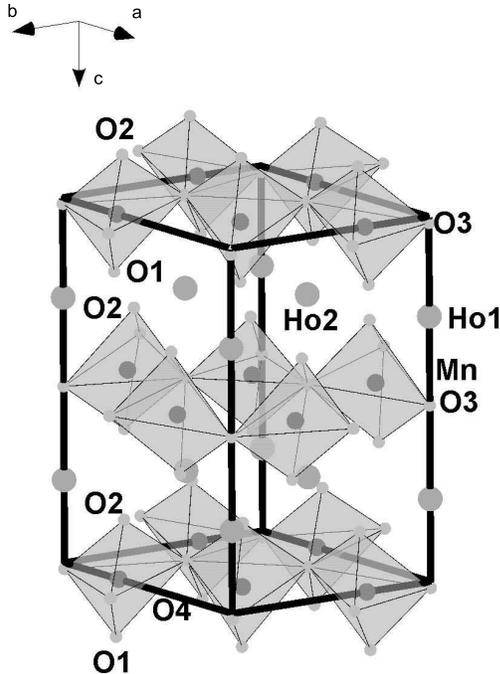}
\caption{Crystallographic structure of hexagonal HoMnO$_3$ (after
parameters of Ref.\cite{munoz1}). }
\end{figure}

The elementary cell of hexagonal $R$MnO$_3$ (space group $P6_3cm$,
C$_{6v}^3$, Z=6) is shown in Fig.~1. The group-theoretical
analysis shows that long-wavelengths zone-center ($\Gamma$-point)
phonons are distributed among irreducible representations of the
C$_{6v}$ point group as follows\cite{iliev2}:

$\Gamma_{tot}$ = 10A$_1$ + 5A$_2$ + 10B$_1$ + 5B$_2$ + 15E$_1$ +
15E$_2$).

\noindent Of them acoustic, Raman-active and infrared-active
phonons are, respectively

$\Gamma_{ac}$ = A$_1$ + E$_1$;

$\Gamma_R$ = 9A$_1$ + 14E$_1$ + 15E$_2$;

$\Gamma_{IR}$ = 9A$_1$ + 14E$_1$.

Optical modes of A$_2$, B$_1$, and B$_2$ symmetries are silent.
A$_1$ vibrations correspond to ion displacements along the $z$
axis. Depending on the scattering configuration one can activate
either longitudinal optical (LO) A$_1$ modes, when phonon
propagation direction coincides with the direction of ion
displacements ($z(xx) \bar{z}$ polarization) or transverse (TO)
modes ($y(xx)\bar{y}$ and $y(zz)\bar{y}$ configurations).
Non-polar E$_2$ vibrations are allowed in $z(xy)\bar{z}$
polarization, while E$_1$ are Raman active in $y(zx)\bar{y}$
polarization. In the latter case experimentally observed E$_1$
modes have the transverse character.

\begin{figure*}
\includegraphics[width=6in]{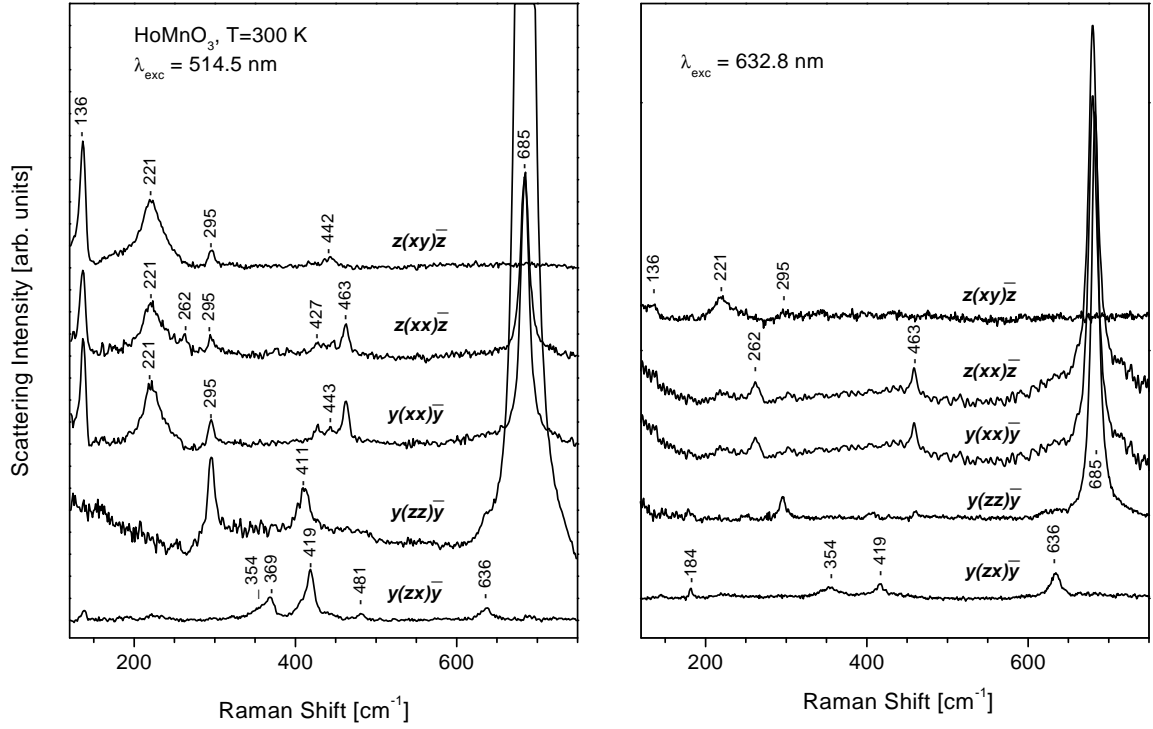}
\caption{Room temperature polarized Raman scattering spectra of
hexagonal HoMnO$_3$ crystals for excitation wavelengths
$\lambda_{exc}$ = 514.5~nm (left panel) and 632.8 nm (right
panel). According to the selection rules, the spectra correspond
to the following symmetries (from top to bottom): E$_2$,
A$_1$(LO)+E$_2$, A$_1$(TO)+E$_2$, A$_1$(TO), and E$_{1}$(TO). }
\end{figure*}

\begin{table*}
\caption{Calculated and experimentally observed at 300K in Raman
spectra $A_1$ and $E_2$ symmetry $\Gamma $-point phonon
frequencies of hexagonal HoMnO$_{3}$. All values are given in
cm$^{-1}$.}
\begin{ruledtabular}
\begin{tabular}{|c|c|c|c|}
\hline
Mode    & Theory  & Experiment   & Direction and sign of \\
        & (TO/LO) & Raman (TO/LO)& the largest atomic displacements \\
\hline
A$_{1}$ & 125/127  &         &  +$z$(Ho/Y1)-$z$(Ho/Y2) \\
A$_{1}$ & 195/234  &         &  rot$_{xy}$(MnO$_{5}$) \\
A$_{1}$ & 245/270  & 262/262 &  +$z$(Ho/Y1,Ho/Y2)-$z$(Mn)\\
A$_{1}$ & 291/295  & 295/295 &  $x$(Mn),$z$(O3) \\
A$_{1}$ & 404/428  & 411/-   &  +$z$(O3)-$z$(O4)+$x,y$(O2)-$x,y$(O1) \\
A$_{1}$ & 430/460  & 427/427 &  +$z$(O4,O3)-$z$(Mn) \\
A$_{1}$ & 468/474  & 463/463   &  +$x,y$(O1,O2)-$x,y$(Mn)\\
A$_{1}$ & 598/614  &         &  +$z$(O1,O2)-$z$(Mn) \\
A$_{1}$ & 673/673  & 685/685 &  +$z$(O1)-$z$(O2) \\
\hline
E$_{2}$ & 64       &      &  $x,y$(Ho/Y1,Ho/Y2,Mn) \\
E$_{2}$ & 96       &      &  +$x,y$(Mn,O3,O4)-$x,y$(Ho/Y1,Ho/Y2) \\
E$_{2}$ & 137      & 136  &  +$x,y$(Ho/Y1)-$x,y$(Ho/Y2) \\
E$_{2}$ & 152      &      &  +$(x,y$(Ho/Y2)-$x,y$(Ho/Y1) \\
E$_{2}$ & 231      & 221  &  +$x,y$(O2,Mn)-$x,y$(O1,O3) \\
E$_{2}$ & 254      &      &  $z$(Mn,O2,O1) \\
E$_{2}$ & 265      &      &  $z$(Mn,O1,O2) \\
E$_{2}$ & 330      & 295  &  +$z$(O2)-$z$(O1),$x,y$(O4) \\
E$_{2}$ & 339      &      &  +$x,y$(O1,O2,O4,O3)-$x,y$(Mn) \\
E$_{2}$ & 402      &      &  +$x,y$(O1,O4)-$x,y$(O2,Mn) \\
E$_{2}$ & 468      & 442  &  +$x,y$(O4)-$x,y$(O1,Mn) \\
E$_{2}$ & 523      &      &  +$x,y$(O4,O3)+$x,y$(O1,O2) \\
E$_{2}$ & 557      &      &  $x,y$(O4) \\
E$_{2}$ & 583      &      &  $x,y$(O4,O3) \\
E$_{2}$ & 649      &      &  $x,y$(O3,O4) \\
\hline
\end{tabular}
\end{ruledtabular}
\end{table*}
\begin{widetext}
\begin{table*}
\caption{Calculated and experimentally observed at 300K E$_1$
symmetry $\Gamma$-point phonon frequencies of hexagonal
HoMnO$_{3}$. All values are given in cm$^{-1}$.}
\begin{ruledtabular}
\begin{tabular}{|c|c|c|c|c|}
\hline
Mode & Theory     & \multicolumn{2}{c}{Experiment} &  Direction and sign of \\
     & TO/LO      & IR (TO/LO) & Raman (TO)        &  the largest atomic displacements \\
\hline
E$_{1}$ & 107/110 &          &                     &  +$x,y$(Mn,O3,O4)-$x,y$(Ho/Y1,Ho/Y2) \\
E$_{1}$ & 143/143 & 136/146  &                     &  +$x,y$(Ho/Y1)-$x,y$(Ho/Y2) \\
E$_{1}$ & 149/149 & 160/164  &                     &  +$x,y$(Ho/Y2)-$x,y$(Ho/Y1) \\
E$_{1}$ & 231/231 &          &                     &  +$x,y$(O1,O2)-$x,y$(Ho/Y1,Ho/Y2) \\
E$_{1}$ & 247/253 & 230/278  &                     &  $x,y$(Mn,O3),$z$(O1,O2) \\
E$_{1}$ & 262/336 & 289/301  &                     &  +$x,y$(O1,O2)-$x,y$(O3) \\
E$_{1}$ & 337/358 & 303/326  &                     &  +$x,y$(O1,O2,O3)-$x,y$(O4,Mn) \\
E$_{1}$ & 359/397 & 355/357  & 354                 &  +$x,y$(O1)-$x,y$(O2) \\
E$_{1}$ & 398/410 & 369/410  & 369                 &  +$x,y$(O1)-$x,y$(O2) \\
E$_{1}$ & 471/491 & 418/476  & 419                 &  +$x,y$(O4,O3)-$x,y$(O2,O1,Mn) \\
E$_{1}$ & 497/537 & 479/551  & 480                 &  +$x,y$(O4,O3,O1,O2)-$x,y$(Mn) \\
E$_{1}$ & 568/571 &          &                     &  $x,y$(O4) \\
E$_{1}$ & 585/586 & 593/601  &                     &  $x,y$(O3) \\
E$_{1}$ & 648/648 &          & 636                 &  $x,y$(O3)-$x,y$(O4) \\
\hline
\end{tabular}
\end{ruledtabular}
\end{table*}
\end{widetext}

As the crystals are non-centrosymmetric, the modes of A$_1$ and
E$_1$ symmetry are also infrared active and modulate the dipole
moment along the $z$-axis and within the $xy$-plane, respectively.
Optical reflection spectra from the large $xy$-plane surfaces
correspond to the case when incident and reflected light propagate
along the $z$-axis and are polarized in the $(xy)$ plane ($\vec{E}
\perp z$). As a consequence, $E_1$ but not A$_1$ phonons are
accessible in this experimental configuration.

Fig.~2 shows the polarized Raman spectra of HoMnO$_3$ at room
temperature obtained with two different excitation wavelengths.
The most intense spectral feature in both cases is the
high-frequency line at 685 cm$^{-1}$, which corresponds to the
$z$-axis apical oxygen (O$_1$, O$_2$) vibrations around the Mn
ions. It is obvious that the relative line intensities depend
strongly on the laser excitation energy. In particular, the line
at 136 cm$^{-1}$ is not observed in the spectra taken with 1.96 eV
($\lambda_{exc}$ = 632.8 nm) excitation. Another striking feature
is the relative intensity variation of the 685 cm$^{-1}$ mode
taken in $(zz)$ and $(xx)$ scattering configuration: for
$\lambda_{exc}$ = 514.5 nm the line intensity in $(zz)$
polarization is by factor of 10 stronger compared to $(xx)$, while
these lines are similar in intensity under $\lambda_{exc}$ = 632.8
nm. All these facts indicate strong resonant contribution to the
Raman scattering spectra, which occurs when incident or scattered
light energies are in resonance with a real electronic transition
in the material under investigation. Indeed, for HoMnO$_3$  a
polarized ($\vec{E} \perp z$) on-side Mn $d-d$ transition near 1.7
eV dominates the absorption spectrum in the visible spectral
range\cite{optcomm,litvin}, similar to the case of
LuMnO$_3$\cite{suchkov}.

\begin{figure}
\includegraphics[width=3in]{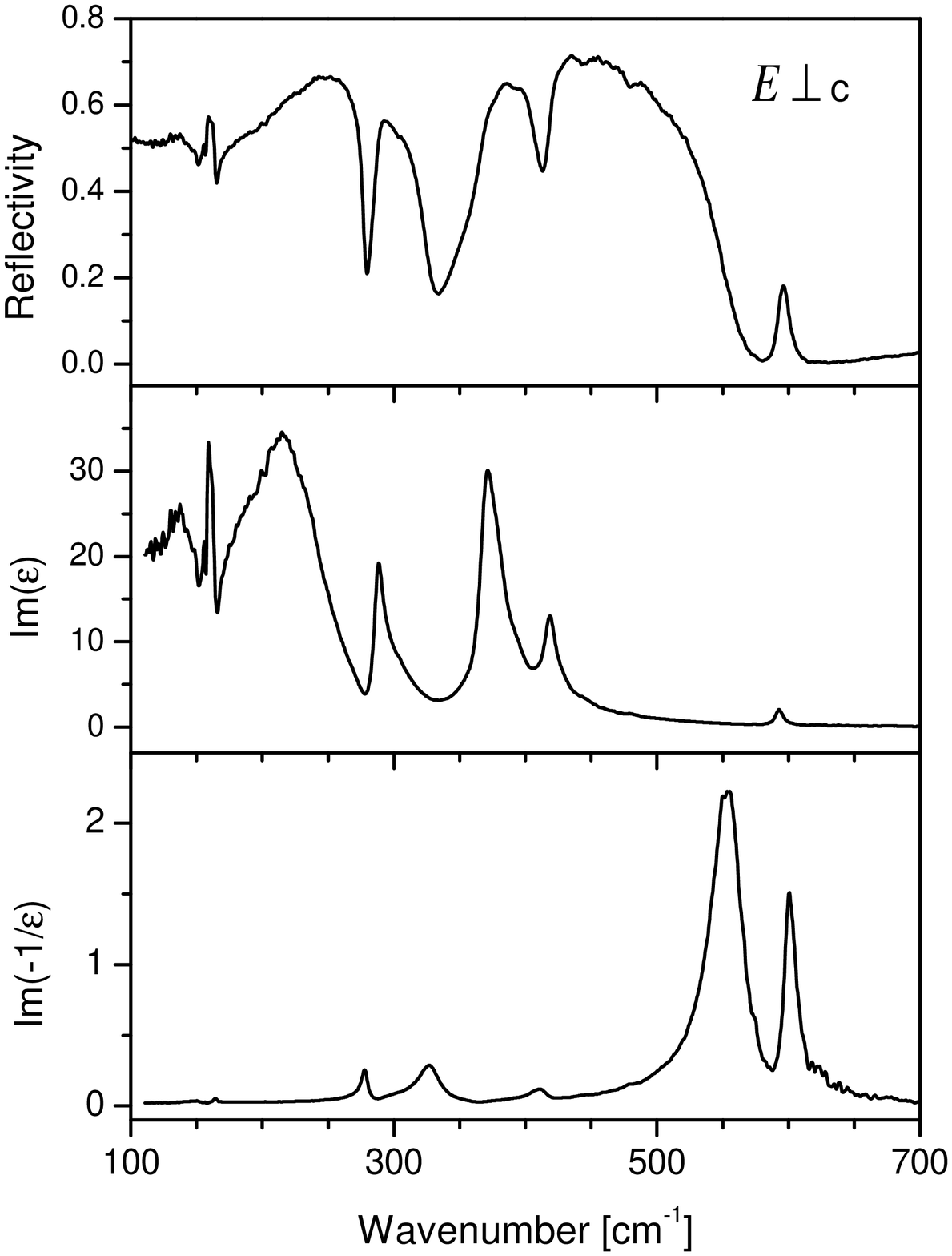}
\caption{Spectral dependence of reflectivity (a), Im($\epsilon$)
(b) and the loss function (c) of HoMnO$_3$ single crystals for
$\vec{E} \perp z$(a) at T=300~K.}
\end{figure}

In Fig.~3 reflection spectra are shown along with the imaginary
part dielectric function $\epsilon_2(\omega)$ and the loss
function Im(-1/$\epsilon$), maxima of which yield the position of
the transverse and longitudinal excitations, respectively. The
experimental and theoretical values of the $\Gamma$-point mode
frequencies are summarized in Tables~I and II.
Similar to the case of isostructural YMnO$_3$
compound\cite{iliev2}, more than half of the expected phonons of
HoMnO$_3$ are observed experimentally and assigned to definite
lattice modes. Comparison of mode frequencies of YMnO$_3$
(Table~IV of Ref.\cite{iliev2}) and HoMnO$_3$ (Table I and II)
shows their close similarity. We also mention that in the case of
E$_1$-symmetry modes, which are both Raman and infrared active,
the experimentally observed TO frequencies of HoMnO$_3$ are in
very good agreement (Table II).
\subsection{Temperature-dependent Raman and infrared spectra}
As far as the temperature dependence of Raman active phonon
parameters is concerned, most of the phonons exhibit standard
anharmonicity-related frequency hardening and linewidth narrowing
upon decreasing temperature, proving the absence of any major
structural transitions in HoMnO$_3$ below 300~K. One of the
Raman-active E$_2$-symmetry phonons is, however, strongly effected
by the antiferromagnetic ordering of Mn ions within $xy$-plane
below T$_N$. As it is seen form the inset in Fig. 4, its frequency
deviates from the dependence expected for the anharmonic decay and
hardens by as much as 5 cm$^{-1}$ between T$_N$ and 10~K.
According to the lattice dynamics calculations (Table I) this
phonon corresponds to the rotations of the MnO$_5$ structural
units in the {\it{xy}}-plane and effectively modulates Mn-O-Mn
exchange interaction. Similar phonon frequency anomalies at the
magnetic ordering temperature have been observed earlier in
CuO\cite{chen95}, SrRuO$_3$\cite{srruo3}, and
CrO$_2$\cite{ilievcr,singcr}.

\begin{figure}
\includegraphics[width=3in]{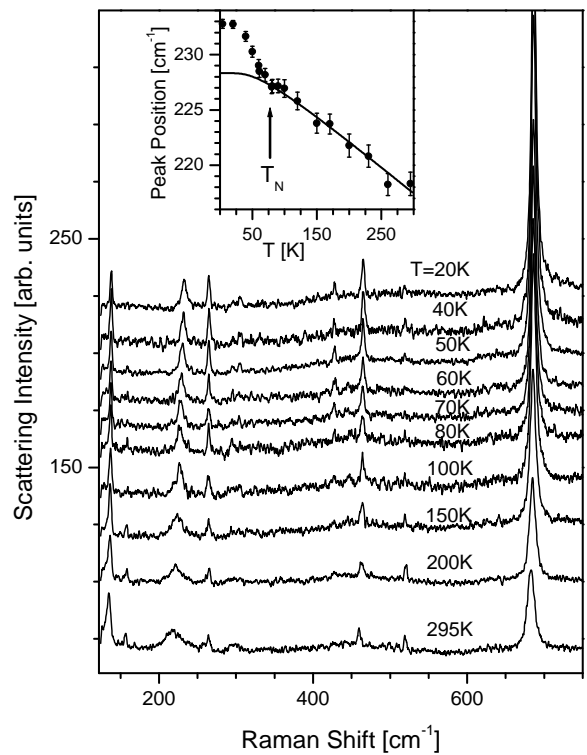}
\caption{Raman scattering spectra of HoMnO$_3$ in $z(xx)\bar{z}$
polarization for various temperatures between 20 and 295~K and
$\lambda_{exc}$ = 514.5~nm. The inset shows the position of the
Mn$O_5$ {\it{xy}}-rotation mode versus temperature. The solid line
shows the peak position expected for a standard
anharmonicity-related phonon decay. T$_N$ marks the N\'eel
temperature.}
\end{figure}

\begin{figure}
\includegraphics[width=3in]{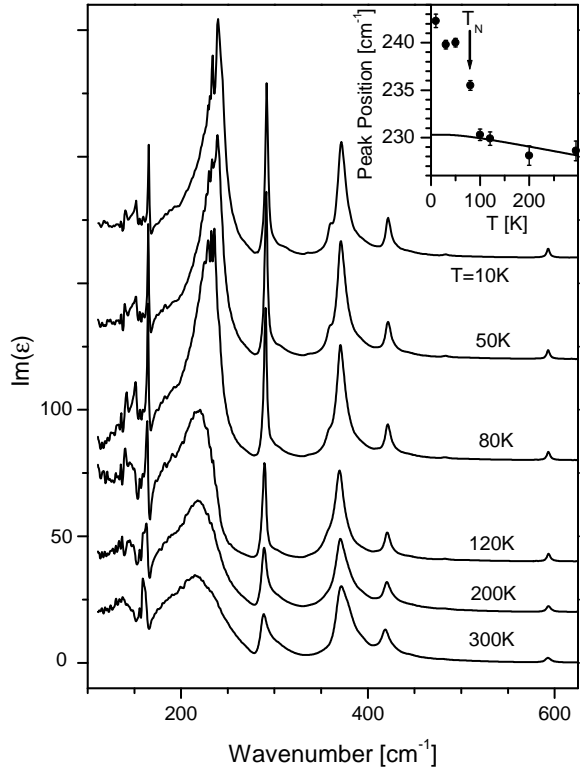}
\caption{Temperature dependence of Im($\epsilon$) for HoMnO$_3$
single crystals ($\vec{E} \perp z$). The spectra are displaced
vertically for clarity. The inset shows the peak position of a
low-frequency mode, which exhibits anomalous hardening below the
N\'eel temperature T$_N$. The solid line is a behavior expected
for a standard anharmonicity-related shift. }
\end{figure}

Another unusual feature, which we noticed in the
temperature-dependent Raman spectra of HoMnO$_3$, is a monotonous
intensity increase (by factor of about three) upon cooling from
300K to 10K of the A$_1$-symmetry line at 685 cm$^{-1}$ in the
$(xx)$ scattering configuration. It may, at least in part, be due
to the temperature-induced shift toward higher energies of the
on-side Mn $d-d$ transition by about 0.19~eV\cite{litvin}, which
affects the resonant conditions.

The spectral dependence of the imaginary part of HoMnO$_3$
dielectric function is shown in Fig. 5 for several temperatures
between 10 and 300~K. Here, like for the Raman-active modes, one
of the modes exhibits remarkable frequency hardening upon entering
the antiferromagnetic state. This E$_1$-symmetry mode, centered at
230 cm$^{-1}$ at room temperature, corresponds primarily to the
in-$(xy)$-plane Mn-O$_3$ bond modulation. Recently similar unusual
behavior of low frequency phonon E$_1$ modes was observed in
isostructural hexagonal LuMnO$_3$ single crystals, suggesting
their coupling to the spin system\cite{suchkov}.
%
Several very narrow lines (the full width at half maximum as low
as 3~cm$^{-1}$), which become especially pronounced at lower
temperatures at around 160~cm$^{-1}$ and 240~cm$^{-1}$, are
related to the transitions between 11 levels of Ho$^{3+}$ ground
multiplet of $^5$I$_8$ symmetry, splitted by the crystalline
electric field. These transitions, as well as those between ground
and excited states, will be discussed in details in a separate
publication\cite{litvin}.

\subsection{Second order Raman scattering spectra
and group-theoretical selection rules}

Unlike the first-order scattering, which due to momentum
conservation probes only zone-center ($\Gamma$-point) vibrations,
the second order scattering involves phonon throughout the entire
Brillouin zone. Scattering intensity is governed in this latter
case by the scattering selection rules and also phonon density of
state, which is determined by the phonon dispersion.

\begin{figure}
\includegraphics[width=3in]{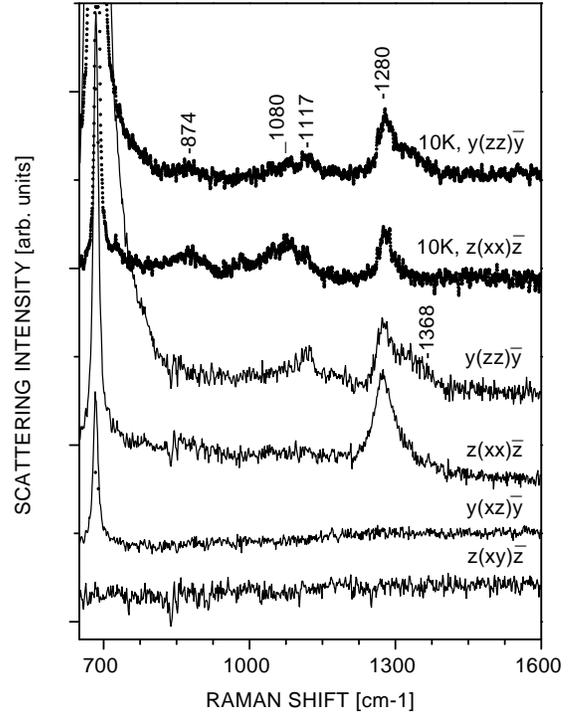}
\caption{Second-order polarized Raman scattering spectra of
HoMnO$_3$ for $\lambda_{exc}$ = 514.5~nm. Two upper spectra are
taken at T=10K, and the lower four at 300K.}
\end{figure}

\begin{table*}
\caption{Group-theoretical selection rules for two-phonon
scattering processes in hexagonal HoMnO$_3$. }
\begin{tabular}{|c|l|l|}
\hline
Special points and      &  &  \\
physically irreducible & Overtones          & Combinations\\
 representations       & & \\
\hline $\Gamma (C_{6v})$  &
[$\Gamma_1$]$^2$,[$\Gamma_2$]$^2$,[$\Gamma_3$]$^2$,[$\Gamma_4$]$^2$$\supset$A$_1$
& $\Gamma_1\times\Gamma_1, \Gamma_2\times\Gamma_2,
\Gamma_3\times\Gamma_3,
\Gamma_4 \times \Gamma_4, \supset$ A$_1$ \\
$\Gamma_1,\Gamma_2,\Gamma_3,\Gamma_4,\Gamma_5,\Gamma_6$ &
[$\Gamma_5$]$^2$, [$\Gamma_6$]$^2$ $\supset$ A$_1$, E$_2$ &
$\Gamma_1\times\Gamma_5, \Gamma_2\times\Gamma_5,
\Gamma_3\times\Gamma_6, \Gamma_4 \times \Gamma_6,
\Gamma_5\times\Gamma_6 \supset$ E$_1$\\
  & & $\Gamma_1 \times \Gamma_6, \Gamma_2\times\Gamma_6,
 \Gamma_3 \times \Gamma_5, \Gamma_4\times\Gamma_5 \supset$ E$_2$ \\
  & & \\
A $(C_{6v})$ & [A$_I]^2$, [A$_{III}]^2 \supset$ A$_1$ &
A$_{III} \times$ A$_{III} \supset$ A$_1$\\
A$_I$=A$_1$+A$_3$ & [$A_{II}]^2$ $\supset$ A$_1$, E$_1$, E$_2$ &
A$_{II} \times$ A$_{II} \supset$
A$_1$, E$_1$, E$_2$\\
A$_{II}$=A$_5$+A$_6$ & & A$_{II} \times$ A$_{III} \supset$ E$_1$,
E$_2$\\
A$_{III}$=A$_2$+A$_4$ & & \\
   & & \\
K $(C_{3v})$ & [$K_1]^2$, [K$_2]^2 \supset$ A$_1$ & K$_1 \times$
K$_1$,
K$_2 \times$ K$_2 \supset$ A$_1$ \\
A$_1$+B$_1 \rightarrow$ K$_1$ & [K$_3]^2 \supset$ A$_1$, E$_1$,
E$_2$ & K$_3 \times$ K$_3 \supset$
A$_1$, E$_1$, E$_2$\\
A$_2$+B$_2 \rightarrow$ K$_2$ &  & K$_1 \times$ K$_3$, K$_2 \times$ K$_3 \supset$ E$_1$, E$_2$ \\
E$_1$+E$_2 \rightarrow$ K$_3$ &  &  \\
   & & \\
M (C$_{2v})$ & [M$_i$]$^2 \supset$ A$_1$, E$_2$ & M$_i \times$ M$_i \supset$ A$_1$, E$_2$ \\
M$_1$, M$_2$, M$_3$, M$_4$ & & M$_1 \times$ M$_2$, M$_3 \times$
M$_4 \supset$ E$_2$ \\
 & & M$_1 \times$ M$_3$, M$_1 \times$ M$_4$, M$_2 \times$ M$_3$,
 M$_2 \times$ M$_4 \supset$ E$_1$ \\
    & & \\
L (C$_{2v})$   & [$L_I]^2$,[L$_{II}]^2\supset$A$_1$,E$_1$,E$_2$ &
L$_I \times$ L$_I$, L$_{II} \times$ L$_{II} \supset$ A$_1$, E$_1$, E$_2$ \\
L$_I$=L$_1$+L$_3$  &  & L$_I \times$ L$_{II} \supset$ E$_1$, E$_2$ \\
L$_{II}$=L$_2$+L$_4$ & & \\
\hline
\end{tabular}
\end{table*}
\begin{table}
\caption{Normal vibrations of HoMnO$_3$ at special points of the
hexagonal Brillouin zone. }
\begin{tabular}{|c|c|}
\hline
Point of the Brillouin    & Irreducible      \\
zone (symmetry)           & representations  \\
\hline $\Gamma$ (C$_{6v}$)  &
5(2$\Gamma_1$+$\Gamma_2$+2$\Gamma_3$+$\Gamma_4$+3$\Gamma_5$+3$\Gamma_6$)\\
$\Delta$ (C$_{6v}$)  &
5(2$\Delta_1$+$\Delta_2$+2$\Delta_3$+$\Delta_4$+3$\Delta_5$+3$\Delta_6$)\\
$A$ (C$_{6v}$)       &
5(2$A_1+A_2+2A_3+A_4+3A_5+3A_6$)\\
    & \\
K (C$_{3v}$)  & 10($2K_1+K_2+3K_3$)\\
P (C$_{3v}$)  & 10($2P_1+P_2+3P_3$)\\
H (C$_{3v}$)  & 10($2H_1+H_2+3H_3$)\\
    & \\
M (C$_{2v}$)   & 5($5M_1+4M_2+5M_3+4M_4$)\\
U (C$_{2v}$)   & 5($5U_1+4U_2+5U_3+4U_4$)\\
L (C$_{2v}$)   & 5($5L_1+4L_2+5L_3+4L_4$)\\
    & \\
$\Sigma$ (C$_s^d$) & 45($\Sigma_1+\Sigma_2$)\\
F (C$_s^d$) & 45($F_1+F_2$)\\
R (C$_s^d$) & 45($R_1+R_2$)\\
    & \\
S (C$_s^v$) & 10(5$S_1+4S_2$)\\
N (C$_s^v$) & 10(5$N_1+4N_2$)\\
T (C$_s^v$) & 10(5$T_1+4T_2$)\\
 \hline
\end{tabular}
\end{table}

The distribution of all 90 normal vibrations of hexagonal
HoMnO$_3$ according to the symmetry type at various points of the
Brillouin zone (see, e.g. \cite{glazer}) is summarized in
Table~III. In this Table special points are grouped according to
their symmetry. In Table~IV the selection rules for overtones and
combination tones are presented for the hexagonal Brillouin zone,
where the notations of Refs.~\cite{gorban,siegle} are used. It
follows from the Table that for all special points of the
Brillouin zone overtones always contributes to the fully symmetric
A$_1$ component, while combinations of phonons belonging to
different symmetries never contain A$_1$ representation.

The second-order scattering spectra of HoMnO$_3$ single crystals
in various scattering configurations and temperatures 300 and 10K
are displayed in Fig. 6. As it is seen, no second-order lines are
detected in the spectra obtained for $(xy)$ and $(xz)$
polarizations (two lowest curves in Fig. 6). The cut-off frequency
of the two-phonon spectrum is expected be at the double frequency
of the highest energy $\Gamma$-point phonon (2 $\times$ 685
cm$^{-1}$ = 1370 cm$^{-1}$). The peak at 1368 cm$^{-1}$, which is
observed in the $(xx)$ polarization, corresponds to the overtone
of the highest frequency A$_1$-symmetry phonon at the Brillouin
zone center. The peak at lower frequency, 1280 cm$^{-1}$, which is
observed in both $(xx)$ and $(zz)$ scattering geometries, is
probably due to the overtone of the same vibration in other
special points (K, M, or L). Under this assumption the estimated
dispersion of the high-frequency branch through the Brillouin zone
is approximately 45 cm$^{-1}$.

    Several peaks in the frequency range 1080-1117, which are
especially pronounced in the $(xx)$ polarization at 10K, are
probably due to combinations of the high-frequency mode with the
phonon branches in the range 350-463 cm$^{-1}$ at various points
of the Brillouin zone. A broad peak centered at 874 cm$^{-1}$ may
be related to the overtones of these latter modes.

We note that the second-order Raman scattering spectra for
hexagonal HoMnO$_3$ are very similar to those reported earlier for
YMnO$_3$ in terms of relative line intensities and position of the
most intense features\cite{iliev1}.

\section{Conclusions}
By means of polarized Raman scattering and infrared reflection
spectroscopy phonon excitations in hexagonal HoMnO$_3$ single
crystals are studied. The zone-center vibrations are assigned to
definite crystal modes based on their polarization properties and
the results of the shell model lattice dynamics calculations. The
results of the group-theoretical analysis of the two-phonon
processes and Raman scattering selection rules are used to
interpret experimental second-order Raman scattering spectra. At
the magnetic transition temperature T$_N$=76K pronounced anomalies
of the Raman- and infrared-active phonon modes, which modulate
Mn-Mn interaction,  are found experimentally and interpreted as
the evidence for spin-phonon coupling.

\acknowledgments This work is supported in part by the State of
Texas through the Texas Center for Superconductivity and Advanced
Materials at the University of Houston. MMG greatly acknowledges
the support of the Bulgarian Science Found (Project F-1207).


\end{document}